\documentclass[11pt,twoside]{article}
\usepackage{asp2014}
\makeatletter
\def\namedlabel#1#2{\begingroup
    #2%
    \def\@currentlabel{#2}%
    \phantomsection\label{#1}\endgroup
}
\makeatother

\aspSuppressVolSlug
\resetcounters

\begin{document}

\title{FAIR standards for astronomical data}

\author{Simon~O'Toole$^1$ (\email{simon.otoole@mq.edu.au})}
\author{James~Tocknell$^1$ (\email{james.tocknell@mq.edu.au})}
\affil{$^1$AAO Macquarie, Macquarie University, NSW, Australia}


\paperauthor{Simon~O'Toole}{simon.otoole@mq.edu.au}{ORCID}{Australian Astronomical Optics}{Author1 Department}{Macquarie University}{NSW}{2113}{Australia}
\paperauthor{James~Tocknell}{james.tocknell@mq.edu.au}{ORCID}{Australian Astronomical Optics}{Author1 Department}{Macquarie University}{NSW}{2113}{Australia}




  
\begin{abstract}

We present an overview of the ``FAIR Guiding Principles for scientific data management and stewardship'', first published in 2016, and how they relate to astronomical data management. In particular, we discuss the connection between the FAIR principles and IVOA standards, and how data management systems with these standards implemented are close to compliance. We then look at what extra components are required to make astronomical data FAIR. Finally, we give a case study of the All-Sky Virtual Observatory (Australia's node of the VO) and their implementation of the FAIR principles.
  
\end{abstract}

\section{Introduction}

It is often said that astronomical data management is "world-leading" and "sets the standard for others to follow". The International Virtual Observatory Alliance (IVOA) is held up (rightly) as an example of how to build a set of standards and protocols for data interoperability and reuse. Now that the FAIR principles are well established, it is appropriate to review and compare how the IVOA standards align with them. This paper attempts to do that, at least at a high level. This work is based on a study funded by the \href{https://ardc.edu.au}{Australian Research Data Commons (ARDC)}.

\section{What are the FAIR Principles?}

The ``FAIR Guiding Principles for scientific data management and stewardship'' were published in 2016 by \citet{2016NatSD...360018W}. They aim to improve the utility of data stores and datasets to researchers employed to analyse scientific data, and implementing them is considered the best practice approach to make data holdings more discoverable and reusable. There are organised under the now well-known four headings: \textbf{F}indable, \textbf{A}ccessible, \textbf{I}nteroperable, and \textbf{R}eusable.

This paper is not intended for a detailed discussion of the FAIR principles, that can be found elsewhere\footnote{E.g. see \href{https://www.go-fair.org/fair-principles/}{https://www.go-fair.org/fair-principles/}}. Here we give a brief overview, which is intended to enable the discussion later in the paper.

\subsection{Findable}\label{sec:findable}
There are four key criteria that make a dataset findable:
\begin{enumerate}
  \item[\namedlabel{item:f1}{F1}] (meta)data are assigned a \emph{globally unique and eternally persistent identifier\\(PID).}
  \item[\namedlabel{item:f2}{F2}] data are described with \emph{rich metadata.}
  \item[\namedlabel{item:f3}{F3}] (meta)data are \emph{registered or indexed in a searchable resource.}
  \item[\namedlabel{item:f4}{F4}] metadata \emph{specify} the data identifier.
\end{enumerate}

\subsection{Accessible}\label{sec:accessible}
A dataset is considered accessible if it satisfies these criteria:
\begin{enumerate}
  \item[\namedlabel{item:a1}{A1}] (meta)data are \emph{retrievable by their identifier} using \emph{a standardized communications protocol.}
  \item[\namedlabel{item:a1p1}{A1.1}] the \emph{protocol} is open, free, and universally implementable.
  \item[\namedlabel{item:a1p2}{A1.2}] the \emph{protocol} allows for an authentication and authorization procedure, where necessary.
  \item[\namedlabel{item:a2}{A2}] \emph{metadata are accessible}, even when the data are no longer available.
\end{enumerate}

\subsection{Interoperable}\label{sec:interoperable}
The three criteria for data interoperability are:
\begin{enumerate}
  \item[\namedlabel{item:i1}{I1}] (meta)data use a \emph{formal, accessible, shared, and broadly applicable language} for knowledge representation.
  \item[\namedlabel{item:i2}{I2}] (meta)data use \emph{vocabularies that follow FAIR principles.}
  \item[\namedlabel{item:i3}{I3}] (meta)data include \emph{qualified references} to other (meta)data.
\end{enumerate}

\subsection{Reusable}\label{sec:reusable}
In order for a dataset to be considered reusable, it must satisfy these criteria:
\begin{enumerate}
  \item[\namedlabel{item:r1}{R1}] meta(data) have a \emph{plurality of accurate and relevant attributes.}
  \item[\namedlabel{item:r1p1}{R1.1}]  (meta)data are released with a \emph{clear and accessible data usage license.} 
  \item[\namedlabel{item:r1p2}{R1.2}] (meta)data are associated with their \emph{provenance.}
  \item[\namedlabel{item:r1p3}{R1.3}] (meta)data \emph{meet domain-relevant community standards.}
\end{enumerate}

\section{Why make your data FAIR?}

For many years, astronomy datasets sat in isolation, requiring users to download all the data and then combine and process them on their laptops. The data was also not always easy to find. This is clearly not efficient, and with the rapidly growing size of data, becoming less and less practical. (The IVOA was set up to address many of these problems.)

Generally speaking, by implementing the FAIR Principles your data becomes easier to find for all researchers. It will also be much easier to combine and integrate with other distributed datasets. This means that your research will have a greater impact on your field. This in turn will make your data more valuable.

Making your data more accessible and easier to find encourages more people (of all ages) to get involved in STEM, even if it's simply a greater appreciation of the benefits of STEM research. The \href{https://zooniverse.org}{Zooniverse} project is a great example of engaging with the general public on all manner of research topics.

\section{FAIR principles in astronomy}

The IVOA develops and maintains standards and protocols which are organised into a coherent architecture; this is outlined in the IVOA Architecture Note \citep{2021ivoa.spec.1101D}. The IVOA Architecture has some similarities with the FAIR Framework, however it focuses on the processes used to move information (e.g. data, metadata, user requests) through the architecture, rather than the properties of the services or data.

\subsection{The IVOA Architecture}

The IVOA Architecture has five components, seen in Figure  \ref{fig:arch2}; these are outlined below.

\articlefigure{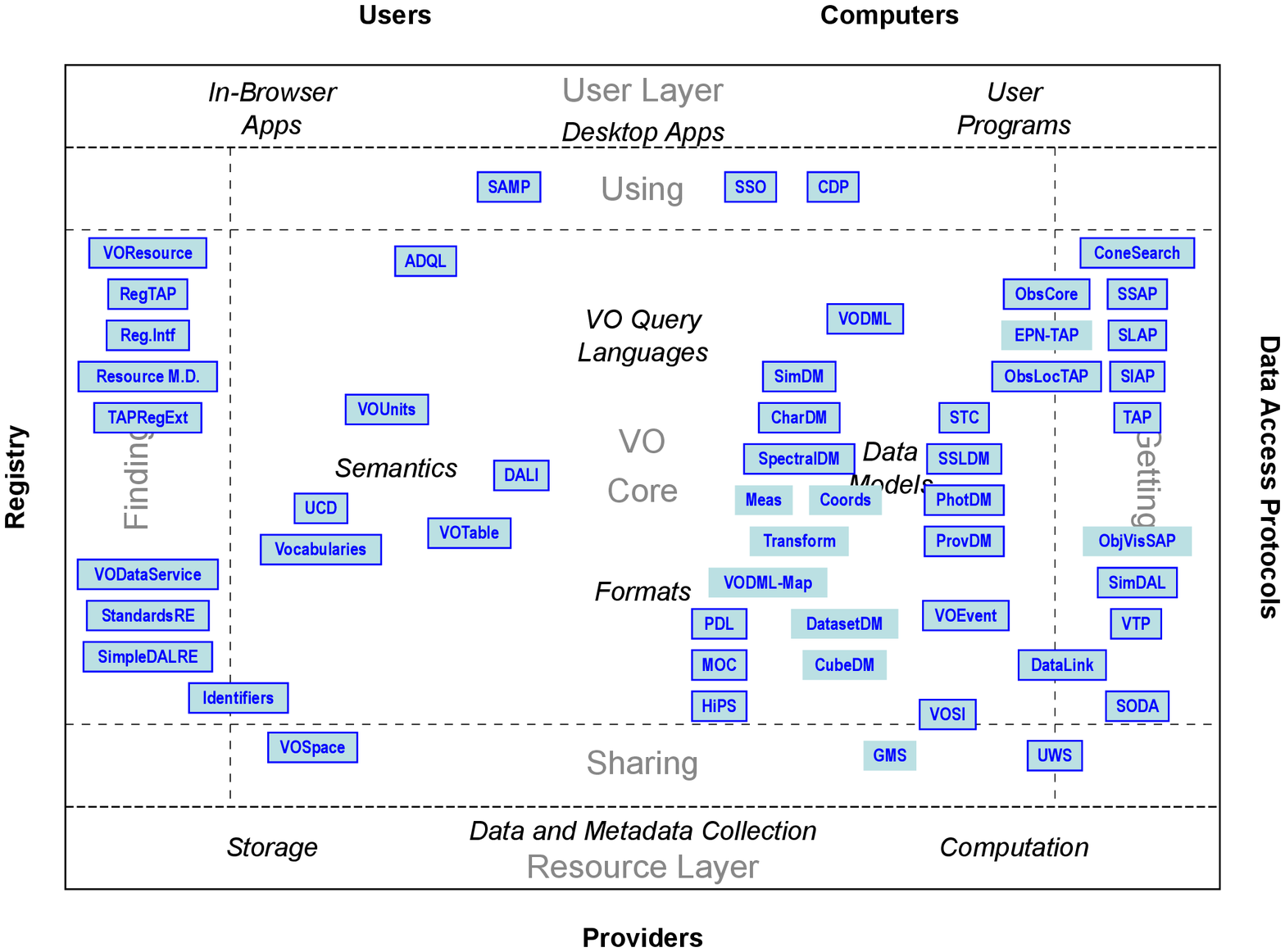}{fig:arch2}{The current IVOA recommendations and in progress recommendations mapped to the five different architecture components.}

\subsubsection{The VO Core component}
This describes how to represent and encode the information transported through a given architecture. IVOA standards within this component include: 
\begin{itemize}
\item data models for:
\begin{itemize}
\item spectra (\href{http://www.ivoa.net/documents/SpectralDM/index.html}{SpectralDM});
\item photometry (\href{http://www.ivoa.net/documents/PHOTDM/index.html}{PHOTDM});
\item spectral line transitions (\href{http://www.ivoa.net/documents/SSLDM/}{SSLDM}); and
\item provenance (\href{http://www.ivoa.net/documents/ProvenanceDM/index.html}{ProvenanceDM});
\end{itemize}
\item the SQL-derived query language \href{http://www.ivoa.net/documents/ADQL/index.html}{ADQL}; and
\item  the \href{http://www.ivoa.net/documents/VOTable/}{VOTable} tabular data format. 
\end{itemize}
  
The component covers most of the \hyperref[sec:interoperable]{Interoperable} and \hyperref[sec:reusable]{Reusable} aspects of FAIR. Notably though, it does not cover \ref{item:r1p1} (licensing), which is node and dataset specific; it is beyond the scope of the IVOA to specify a licensing regime for data. The licensing question is discussed below.

\subsubsection{The Finding component (a.k.a. Registry)} 
This maps nicely to the \hyperref[sec:findable]{Findable} aspect of FAIR. It covers the metadata and its encoding, for both the data within the VO service, and the VO service itself. Standards with this component include \href{http://www.ivoa.net/documents/IVOAIdentifiers/20160523/REC-Identifiers-2.0.html}{IVOA Identifiers}, which provides a global identifier as required by \ref{item:f1}.
  
\subsubsection{The Getting component (a.k.a Data Access Protocols)} 
This contains the different protocols used to access and download the diversity of data and metadata that is contained within astronomical datasets. It covers such protocols such as: 
  \begin{itemize}
      \item the \href{http://www.ivoa.net/documents/TAP/}{Table Access Protocol} (TAP) which allows accessing and downloading both data and metadata within VO service, and performing additional operations on this data (such as filtering, joining and summarising) via ADQL and other query languages; 
      \item the \href{http://www.ivoa.net/documents/latest/ConeSearch.html}{Simple Cone Search} protocol, which allows finding astronomical objects within a specified angular radii from a point; 
      \item the \href{http://www.ivoa.net/documents/SIA/}{Simple Image Access} protocol (SIA), which has a similar purpose to TAP, but with the proviso that it works on imaging data (both 2D and higher), rather than tabular data; 
      \item the \href{http://www.ivoa.net/documents/SLAP/index.html}{Simple Line Access Protocol} (SLAP) for accessing spectral line information; and 
      \item the \href{http://www.ivoa.net/documents/SSA/}{Simple Spectral Access} protocol (SSA) for accessing 1D spectra. 
  \end{itemize}
 
These protocols map to the \ref{item:a1} \hyperref[sec:accessible]{Accessible} aspect of FAIR. \ref{item:a2} (metadata are accessible even when data is gone) is node and dataset specific, and will require each node to provide some kind of policy around the storage of metadata in the long term.

\subsubsection{The Using component} 
This contains specific user-facing aspects, namely the integration between applications and access control. It covers protocols including: 
  \begin{itemize}
      \item the \href{http://www.ivoa.net/documents/SAMP/index.html}{Simple Application Messaging Protocol} (SAMP), which allows data to be passed between applications;
      \item \href{http://www.ivoa.net/documents/SSO/index.html}{Single Sign On} (SSO), which covers the authorisation and authentication protocols needed to access resources; and 
      \item \href{http://www.ivoa.net/documents/CredentialDelegation/}{Credential Delegation Protocol} (CDP), which allows VO services to make requests on behalf of users.
  \end{itemize}
  
SSO and CDP provide the authentication and authorisation component of \ref{item:a1p2} to the Data Access Protocols mentioned above. SAMP does not naturally fit under any of the FAIR principles (the closest being \ref{item:a1}), as it focuses on interoperability between applications, rather than the actual data.
  
\subsubsection{The Sharing component} 
This focuses on the provision of computation and storage as needed to use the other VO standards and protocols. TAP for example may use the Universal Worker Service (\href{http://www.ivoa.net/documents/UWS/index.html}{UWS}) pattern for running a complex query that would take a non-trivial time to complete. Like SAMP, these protocols and standards do not naturally fit under any of the FAIR principles, given their process-orientated nature.

\subsection{IVOA Standards and FAIR}
\label{sec:ivoa-fair}
Given how many of the FAIR principles are covered by the IVOA standards and protocols, services which correctly implement and use these standards and protocols are quite close to providing a fully-FAIR resource. However there are a few specific principles which the IVOA standards and protocols do not provide, generally because they are out of scope. This paper therefore focuses on these specific principles.

\subsubsection{IVOA standards are (mainly) Findable}

Our study has found that there are two criteria in the \hyperref[sec:findable]{Findable} section of the FAIR principles that IVOA standards do not address: the assignment of an eternally persistent identifier (\ref{item:f1}), and that the data identifier is returned in all queries (\ref{item:f4}). 

A globally unique identifier is part of the IVOA standards through its IVOID, as discussed above, however there is currently no recommendation that requires it to be eternally persistent. This requirement deals with the situation if/when systems are transitioned (e.g. how are identifiers preserved/mapped across database restructurings), or when organisations are restructured (so URLs are forced to change) or when the funding runs out. It requires the nodes have policies to handle these transitions or events; we discuss how implementations might address this below. 

Note that IVOIDs do not currently appear to be citable in the same way as Digital Object Identifiers (DOIs), as would be preferred by e.g. the \href{https://ardc.edu.au/resources/working-with-data/fair-data/fair-self-assessment-tool/}{ARDC FAIR self-assessment tool}. This is a natural consequence of IVOIDs being a VO-specific identifier, rather than a global identifier used to manage citations.

IVOA standards do not require that the data identifier be returned in all cases, and most FITS files will contain no data-specific identifier. Satisfying \ref{item:f4} would require each node to check code paths to ensure that what data-specific identifiers are in the data are correctly propagated (and not discarded along the way), ensure that data identifiers are created or correctly ingested for both existing and incoming data, and that existing services always provide the data identifier(s). The IVOA standards leave this principle to service providers to decide what to do.

\subsubsection{IVOA standards are (mainly) Accessible}

The FAIR principle \ref{item:a2} above requires that the metadata of a dataset be accessible even if the data are no longer available. This goes to the question of service reliability. The IVOA standards do not specify any rules about how reliable a service must be, so this is entirely up to those implementing them as to how to handle problems where the data is no longer available. 

There are several validation services available for the IVOA standards in the \textit{Getting} component of the architecture, which help determine the availability of services. These are then used by the IVOA \href{https://wiki.ivoa.net/twiki/bin/view/IVOA/IvoaOps}{Operations Interest Group} to generate ``weather reports'' that are presented periodically at IVOA meetings. We note that these validation services are useful for standards compliance, and simply \textit{report} whether a system and its data/metadata is available or not; they do not by themselves satisfy \ref{item:a2}.

\subsubsection{IVOA standards are (pretty much) Interoperable}

The IVOA Standards implement vocabularies and in particular Universal Content Descriptors that mean both \ref{item:i1} and \ref{item:i2} are fully met. The third principle \ref{item:i3} in \hyperref[sec:interoperable]{Interoperability} requires ``(meta)data to include qualified references to other (meta)data'', and here the situation is less clear. The description of this principle
from  \href{https://www.go-fair.org/fair-principles/i3-metadata-include-qualified-references-metadata/}{GO-FAIR} organisation is: ``A qualified reference is a cross-reference that explains its intent''. GO-FAIR recommends using the Resource Description Framework (RDF) to satisfy this requirement, however they do not explain how to use it to provide the meaningful qualified references. 

The IVOA \href{https://www.ivoa.net/documents/DataLink/20150617/index.html}{DataLink} standard describes "the linking of data discovery metadata to access to the data itself, further detailed metadata, related resources, and to services that perform operations on the data." This therefore appears to satisfy \ref{item:i3}, although often when the DataLink standard is implemented, the references are for different types of data that describe the same object in a single dataset, rather than across different datasets.

We note that typically the most common type of cross-referencing done in astronomy between different datasets are mappings between different identifiers of the same astronomical object. \href{https://datacentral.org.au}{Data Central}, for example, provides a Name Resolver service, which searches through multiple databases to try to find all the names (or identifiers) the object has been given. Surveys, especially ones which observe previously surveyed areas of the sky, try now to provides a mapping between their survey identifier and previous survey identifiers, avoiding the need for astronomers to manually cross-match objects in different datasets. 

One could therefore argue that \textit{all} of the \hyperref[sec:interoperable]{Interoperability} principles are met, at least to some degree.

\subsubsection{IVOA standards are (almost) Re-usable}

The key issue that the IVOA standards do not address is that around data licenses (\ref{item:r1p1}). The IVOA has no specifications about what license that data and metadata should be released under, and it is left to providers or data producers to determine the best licenses to use. Many funding bodies (e.g. Australian Research Data Commons) are requesting that groups apply licenses to data releases, so many datasets are now being released with licenses, or having them retrospectively applied. The \href{https://creativecommons.org/}{Creative Commons} set of licenses are becoming a \textit{de facto} standard in this area.

When we conducted our original study of the FAIRness of IVOA standards in 2019, data provenance was still something not covered. The IVOA \href{https://www.ivoa.net/documents/ProvenanceDM/20200411/index.html}{Provenance Data Model} was published in 2020, so the principle addressing this (\ref{item:r1p2}) is covered. We note though, that there will be a significant period of time before the data model is deployed across different sites.

\section{A Case Study: the All-Sky Virtual Observatory}

The main driver of our original FAIRness study was to examine how the Australian All-Sky Virtual Observatory (ASVO) implemented FAIR principles and what areas still need improvement.
The five nodes in the ASVO are described in \citet{2019ASPC..523..413O}. Briefly, they are: the \href{https://casda.asvo.org.au}{CSIRO ASKAP Science Data Archive (CASDA)}; \href{https://datacentral.asvo.org.au}{Data Central}, which includes the Anglo-Australian Telescope archive and other high-level science products; the \href{https://mwa.asvo.org.au}{Murchison Widefield Array (MWA) ASVO}; the \href{https://skymapper.asvo.org.au}{SkyMapper data archive}; and the \href{https://tao.asvo.org.au}{Theoretical Astrophysical Observatory (TAO)}, which hosts data from multiple popular cosmological simulations and galaxy formation models, and allows users to build custom mock galaxy catalogues and images.

The ASVO nodes' use the IVOA protocols and standards, meaning that the FAIRness or otherwise of these standards and protocols in large part governs the FAIRness of the nodes both individually and as a collective. In late 2018, the ASVO nodes were asked to assess how FAIR they believed they were; the results are presented in Table \ref{first-fair-table}.

\begin{table}
  \centering
\begin{tabular}{llllll}
\hline

& CASDA & Data Central & MWA & Skymapper & TAO\tabularnewline

\hline

F1 & Y & Y & Y & Y & Y\tabularnewline
F2 & Y & Y & Y & Y & Y\tabularnewline
F3 & Y & Y & Y & Y & Y\tabularnewline
F4 & P & P & P & P & P\tabularnewline
A1 & Y & Y & Y & Y & Y\tabularnewline
A1.1 & Y & Y & Y & Y & Y\tabularnewline
A1.2 & Y & Y & Y & Y & Y\tabularnewline
A2 & Y & Y & Y & Y & Y\tabularnewline
I1 & Y & Y & Y & Y & Y\tabularnewline
I2 & P & Y & Y & Y & P\tabularnewline
I3 & P & P & P & P & P\tabularnewline
R1 & Y & Y & Y & Y & Y\tabularnewline
R1.1 & Y & P & P & P & Y\tabularnewline
R1.2 & P & P & P & P & P\tabularnewline
R1.3 & Y & Y & Y & Y & Y\tabularnewline

\hline

\end{tabular}
\caption{Estimated level of FAIRness of the different ASVO nodes: Y is Yes, P is
  partial, in development or needs improvement.\label{first-fair-table}}
\end{table}

It is worth noting that each of the nodes provides a different type of output, and some of these outputs are more amenable to be made FAIR than others, whether due to the age of the data (meaning that information/knowledge has been lost or not documented), whether the data is generated or recalled from storage (and hence how to mint identifiers), or the level of modification needed to use the data (which also affects identifiers). We also point out that while the TAO node satisfies many of the FAIR principles, it does so \textit{without} using IVOA standards, since at the beginning of its development, there were no standards designed to manage theoretical data.

\subsection{CASDA}
CASDA addresses most of the FAIR principles including the non-IVOA principles above. For example:
\begin{itemize}
    \item \ref{item:f1} CASDA mints DOIs and IVOIDs, both at the dataset level; an example can be found at \href{https://doi.org/10.25919/5d3c3dc31999d}{https://doi.org/10.25919/5d3c3dc31999d}
    \item \ref{item:a2} This principle is addressed by policies such as the CSIRO \href{https://confluence.csiro.au/display/daphelp/Data+Preservation+Principles}{Data Presevation Principles}, which form part of the CSIRO \href{https://www.coretrustseal.org/}{CoreTrustSeal} award.
    \item \ref{item:r1p1} All CASDA data is release with a license, with licenses listed on the \href{https://confluence.csiro.au/display/dap/Licence+deeds}{Data Access Portal license page}.
    \item \ref{item:r1p2} Provenance details such as as observing dates, Project IDs, Scheduling Block IDs and software version, are stored in a database and made accessible through the CASDA TAP service.
\end{itemize}

\subsection{Data Central}
Data Central addresses most of the FAIR principles and some of the non-IVOA principles discussed above, including:
\begin{itemize}
    \item \ref{item:f1} DOIs are not currently minted, however identifiers are used on a per-survey basis, which while not globally unique, can be dereferenced via the association with their respective survey.
    \item \ref{item:a2} Data Central is looking at developing policies to handle the preservation of metadata in the case where the service is shut down (e.g. due to loss of funding).
    \item \ref{item:r1p1} Currently Data Central does not provide a license under which the all of data is provided, however is planning to provide access to new survey data under the Creative Commons Attribution license.
    \item \ref{item:r1p2} Data provenance information can be provided by the research teams that host their data in Data Central through various means including a schema browser and documentation portal
\end{itemize}

\subsection{MWA ASVO}
MWA ASVO addresses most of the FAIR principles, including the non-IVOA principles discussed above. Some issues include:
\begin{itemize}
    \item \ref{item:f1} MWA ASVO hosts raw data only, so minting DOIs is problematic. The MWA TAP service provides a globally unique identifier through an IVOID. 
    \item \ref{item:a2} As with many SKA pathfinder datasets, long term data hosting is guaranteed until at least 2027, although there is currently no specific policy around what would happen if the data were no longer available.
    \item \ref{item:r1p1} The MWA Collaboration has a \href{https://www.mwatelescope.org/images/documents/2012.08.15_MWA_data_access_policy_approved.pdf?type=file}{Data Access Policy} which details how the data may be used within and outside the Collaboration. This closely matches the \href{https://creativecommons.org/licenses/by-nc/4.0/}{Attribution-Non Commercial 4.0 International Creative Commons license}.
    \item \ref{item:r1p2} MWA stores provenance metadata in their TAP service.
\end{itemize}

\subsection{SkyMapper}
SkyMapper addresses the FAIR principles including most of the non-IVOA principles discussed above. For example:
\begin{itemize}
    \item \ref{item:f1} Each SkyMapper data release has its own DOI minted.
    \item \ref{item:a2} SkyMapper plans to host the SkyMapper website and related metadata on Mount Stromlo Observatory systems indefinitely.
    \item \ref{item:r1p1} SkyMapper data is currently not released to the public under a specific license.
    \item \ref{item:r1p2} Provenance information is provided by the SkyMapper \href{https://skymapper.anu.edu.au/table-browser/}{Table Browser}.
\end{itemize}

\subsection{TAO}
As mentioned above, TAO does not implement many IVOA standards at all, given there are few available in the theory/simulation space at this time. This has an impact in particular on interoperability as can be seen in Table \ref{first-fair-table}. Despite this, TAO addresses many of the FAIR principles, including the non-IVOA principles discussed above, including:
\begin{itemize}
    \item \ref{item:f1} As the data is described by the simulation input parameters, having the original inputs is sufficient to describe and recreate the data, and users can a request that a DOI be minted for any dataset.
    \item \ref{item:a2} As the data and metadata can be generated from the input parameters, and the code to do so is publicly available, it should always be possible to recreate the data from TAO, if it were to disappear. TAO is also a distributed system across multiple supercomputing centres, which leads to higher availability.
    \item \ref{item:r1p1} TAO does not currently have a publicly available data license.
    \item \ref{item:r1p2} Provenance data is included with all generated TAO datasets.
\end{itemize}

\section{Key takeaways}

We have presented a overview of how the IVOA standards map to the FAIR principles, with a focus on which principles the IVOA standards do not cover. The key message is: if you implement IVOA standards and protocols to provide access to your data, then your data will be \textit{almost} FAIR! As we have discussed in Section \ref{sec:ivoa-fair}, with just a few extra additions you can make your data completely FAIR. In effect, the following three steps are enough to make your data FAIR:
\begin{enumerate}
    \item Use IVOA standards and protocols
    \item Mint DOIs
    \item Add a license
\end{enumerate}

There are now multiple implementations of the key IVOA standards discussed in this paper, so it is easier than ever to make your data VO-compliant!

\bibliography{I4-001}

\begin{thebibliography}{}
\expandafter\ifx\csname natexlab\endcsname\relax\def\natexlab#1{#1}\fi
\expandafter\ifx\csname url\endcsname\relax
  \def\url#1{\texttt{#1}}\fi
\expandafter\ifx\csname urlprefix\endcsname\relax\def\urlprefix{URL }\fi
\providecommand{\eprint}[2][]{\url{#2}}

\bibitem[{{Dowler} et~al.(2021){Dowler}, {Evans}, {Arviset}, {Gaudet}, \&
  {Technical Coordination Group}}]{2021ivoa.spec.1101D}
{Dowler}, P., {Evans}, J., {Arviset}, C., {Gaudet}, S., \& {Technical
  Coordination Group} 2021, {IVOA Architecture Version 2.0}, IVOA Endorsed Note
  01 November 2021

\bibitem[{{O'Toole} \& {Sealey}(2019)}]{2019ASPC..523..413O}
{O'Toole}, S., \& {Sealey}, K. 2019, in Astronomical Data Analysis Software and
  Systems XXVII, edited by P.~J. {Teuben}, M.~W. {Pound}, B.~A. {Thomas}, \&
  E.~M. {Warner}, vol. 523 of Astronomical Society of the Pacific Conference
  Series, 413

\bibitem[{{Wilkinson} et~al.(2016){Wilkinson}, {Dumontier}, {Aalbersberg},
  {Appleton}, {Axton}, {Baak}, {Blomberg}, {Boiten}, {da Silva Santos},
  {Bourne}, {Bouwman}, {Brookes}, {Clark}, {Crosas}, {Dillo}, {Dumon},
  {Edmunds}, {Evelo}, {Finkers}, {Gonzalez-Beltran}, {Gray}, {Groth}, {Goble},
  {Grethe}, {Heringa}, {'T Hoen}, {Hooft}, {Kuhn}, {Kok}, {Kok}, {Lusher},
  {Martone}, {Mons}, {Packer}, {Persson}, {Rocca-Serra}, {Roos}, {van Schaik},
  {Sansone}, {Schultes}, {Sengstag}, {Slater}, {Strawn}, {Swertz}, {Thompson},
  {van der Lei}, {van Mulligen}, {Velterop}, {Waagmeester}, {Wittenburg},
  {Wolstencroft}, {Zhao}, \& {Mons}}]{2016NatSD...360018W}
{Wilkinson}, M.~D., {Dumontier}, M., {Aalbersberg}, I.~J., {Appleton}, G.,
  {Axton}, M., {Baak}, A., {Blomberg}, N., {Boiten}, J.-W., {da Silva Santos},
  L.~B., {Bourne}, P.~E., {Bouwman}, J., {Brookes}, A.~J., {Clark}, T.,
  {Crosas}, M., {Dillo}, I., {Dumon}, O., {Edmunds}, S., {Evelo}, C.~T.,
  {Finkers}, R., {Gonzalez-Beltran}, A., {Gray}, A. J.~G., {Groth}, P.,
  {Goble}, C., {Grethe}, J.~S., {Heringa}, J., {'T Hoen}, P. A.~C., {Hooft},
  R., {Kuhn}, T., {Kok}, R., {Kok}, J., {Lusher}, S.~J., {Martone}, M.~E.,
  {Mons}, A., {Packer}, A.~L., {Persson}, B., {Rocca-Serra}, P., {Roos}, M.,
  {van Schaik}, R., {Sansone}, S.-A., {Schultes}, E., {Sengstag}, T., {Slater},
  T., {Strawn}, G., {Swertz}, M.~A., {Thompson}, M., {van der Lei}, J., {van
  Mulligen}, E., {Velterop}, J., {Waagmeester}, A., {Wittenburg}, P.,
  {Wolstencroft}, K., {Zhao}, J., \& {Mons}, B. 2016, Scientific Data, 3,
  160018

\end{thebibliography}

\end{document}